\documentclass[aps,preprint,preprintnumbers,amsmath,amssymb]{revtex4}
\usepackage{amsmath,mathrsfs,amsbsy,color,graphicx,bm,amsthm,amsfonts}
\usepackage{units}
\usepackage{bbm}
\usepackage{times}
\usepackage{dcolumn}
\usepackage{mathrsfs}
\usepackage{amsmath,amssymb,epsfig}
\usepackage{hyperref}

\begin{document}

\title{ Entanglement-enhanced quantum ranging in the near-Earth spacetime}

\author{Qianqian Liu$^{1}$, Cuihong Wen$^{1}$, Jiliang Jing$^{1}$, and Jieci Wang$^{1}$\footnote{Email: jcwang@hunnu.edu.cn}}
\affiliation{$^1$  Department of Physics and Synergetic Innovation Center for Quantum Effects,\\Key Laboratory of Low-Dimensional Quantum Structures and Quantum Control of Ministry of Education, \\
Key Laboratory for Matter Microstructure and Function of Hunan Province,\\
 Hunan Normal University, Changsha 410081, China}

\begin{abstract}
 We propose a quantum ranging protocol to determine the distance between an observer and a target at the line of sight in the near-Earth curved spacetime.
 Unlike the quantum illumination scheme, here we employ multiple quantum hypothesis testing to decide the presence and location of the target at the same time.
  In the present protocol, the gravity of the Earth influences the propagation of photons and  the performance of  quantum ranging.
We find that the maximum potential advantages of the quantum ranging strategy in the curved spacetime outperform its flat spacetime counterpart.
This is because the effect of gravitational red-shift and blue-shift on the  entangled signal photons can be canceled out, while the thermal photons only suffers from the gravitational  blue-shift effect.
We also show that the number of transmitted modes can promote the maximum potential advantage of the quantum ranging tasks. The maximum potential advantage of quantum ranging in the curved spacetime can not be raised sharply by dividing the range into multiple slices.
\end{abstract}
\vspace*{0.5cm}

\maketitle
\section{Introduction}\label{section1}

 Quantum illumination (QI) \cite{Lloyd,Tan,Zhuang1,Zhuang2,Zhuang3,Zhuang4} is an entanglement-assisted target detection scheme that uses an optimal quantum receiver and has a good advantage over the optimal classical strategy in terms of error exponent.
 In addition, the employment of entanglement can solve the disadvantage of the rapid attenuation of traditional radar signals.
 Recently, many efforts have been put forward to  make the QI schemes' theoretical advantage practically relevant \cite{pract1,pract2,pract3},  and some of these schemes have been recently experimentally demonstrated  \cite{expriment1,expriment2,expriment3}. However,  the extension of the QI scheme
 can only query a single spatiotemporal resolution bin at a time, which limits the realization of quantum radar in the real world \cite{hypoth4}.
Fortunately, quantum ranging protocols have solved this limitation in recent work \cite{ranging1}.
In such a proposal, the transmitter sends signal pulses to the target region and performs continuous measurements at the receiver side to determine the reflection of the target at the line of sight \cite{ranging1,ranging2}.  The main advantage of quantum ranging  is  the employment of  the multiple quantum hypothesis testing scenario \cite{hypoth1,hypoth2,hypoth3} instead of the  binary hypothesis  to determine the existence and location of the target at once. The quantum ranging task is formulated as a multiple hypothesis testing problem \cite{ranging1}, where each hypothesis corresponds to the conclusion that the target exists in some distance slice. The observer on the ground transmit signal pulses to the target region. We assume that the distance can be divided  into some finite discrete intervals, and the hypotheses are assigned corresponding to one of the range slices.  Then the task of ranging is realized by the determination of the reflected mode among the continuously collections. For this reason, the performance of the ranging strategy is determined by the error probability of the hypotheses testing.

 On the other hand, the novel field of relativistic quantum information aims to understand the preparation, manipulation, and transmission of quantum information in a relativistic setting \cite{re1,re2,re3,RQI1,RQI2, RQI3, gdecoheren1,gdecoheren2,vis,tabletop,clock,earthqs1,Rquan,earthqs4}.
 In particular, many experimental and theoretical proposals have been put forward to measure gravity induced decoherence of a quantum state \cite{gdecoheren1,gdecoheren2}, and test the quantum nature of gravity with tabletop experiments \cite{tabletop} based on quantum entanglement properties. One can use the  properties of quantum entanglement to better understand quantum clocks within relativistic settings \cite{ Rquan}, and investigate how the gravitational field of the Earth affect satellite based quantum communication \cite{earthqs1,earthqs4} and  clock synchronization  \cite{clock}.   It is widely believed that the study of  quantum information within the framework of relativity can provide new insights into some basic questions in quantum mechanics and relativity, including nonlocality, causality and the information paradox of black holes. More importantly, it is of practical significance to clarify the roles of relativistic effects in realistic quantum information tasks when the parties are separated by far in the near-Earth curved spacetime.

Quantum entanglement has been demonstrated to play a key role in the quantum ranging protocol \cite{ranging1}.
Entanglement is a fragile quantum resource that is easily corrupted by noise and loss.
 It is worth pointing out that the propagation pulse is affected via changing their frequency distribution in center and shape in the near-Earth curved spacetime \cite{earthqs1,earthqs4,earthqs2,earthqs5,earthqs6,clock}. The Earth's gravity has been found to make observable effects on the entanglement and  fidelity of quantum communication \cite{earthqs1,earthqs4},  the precision of quantum metrology \cite{earthqs2,earthqs5,earthqs6}, and the reliability of  quantum clock synchronization  \cite{clock}. Therefore, the gravitational field of the Earth should be taken into serious consideration for practical quantum ranging tasks.

In this paper, we propose a quantum ranging protocol in a more realistic scenario where the nonmaskable gravity of the Earth is considered. We are interested in  how the Earth's gravity influences the detection performance of the quantum ranging protocol.  It is assumed that one component of  the entangled signal-idler photon pair is sent from the Earth to a spatial target region  to perform the quantum ranging task. The wave packet overlap and the transmissivity of the photons are deformed by the Earth's spacetime curvature during the propagation. The maximum potential quantum advantage  \cite{thermal}  is employed to represent the maximum possible improvement of quantum ranging protocol compared with classical ranging.  It is shown that the maximum potential advantages of  the quantum  ranging  strategy in the curved spacetime has distinct superiority over its counterpart in the flat spacetime.


 This paper is organized as follows. In Sec. II, we introduce the propagation of the photons under the background of the Earth. In Sec. III, we briefly introduce the quantum ranging tasks and calculate the potential maximum quantum advantage in different spacetime.
Finally, the conclusions are drawn in Sec. IV.

\section{Light wave packets propagating in Earth}\label{section2}

In this section, we discuss the transmission of light wave packets from the ground to the target region \cite{earthqs1,earthqs2,earthqs4}.
 The Earth's spacetime can be approximately described by the Kerr metric,  which well approximates the rotating spherical planet.
We restricted this paper to the equatorial plane $\theta=\frac{\pi}{2}$ and get the reduced Kerr metric in the Boyer-Lindquist coordinates $(t,r,\phi)$ \cite{kerr}
\begin{eqnarray}\label{metric}
ds^2&=&\, -\Big(1-\frac{2M}{r} \Big)dt^2+\frac{1}{\Delta}dr^2+\Big(r^2+a^2+\frac{2Ma^2}{r}\Big) d\phi^2 - \frac{4Ma}{r} dt \, d\phi,
\end{eqnarray}
where $\Delta =1-\frac{2M}{r}+\frac{a^2}{r^2}$, Kerr parameter $a=\frac{J}{M_{A}}$, and $r_{A}$, $M_{A}$,  $J$ are the radius, mass, angular momentum of the Earth. Throughout this paper we set
$\hbar=c=1$.

We only consider the  radial propagation case. It is reasonable because the angular velocity of the Earth is negligible and the frequencies of the observers on  the Earth are too small compared to the characteristic frequencies involved. In this case, the evolution of the quantum field is a 1+1 dimensional problem. Since the uncharged scalar field is a good approximation of the Maxwell electromagnetic field in longitudinal or transverse modes \cite{RQI3,Downes}, we can restrict our analysis based on the solutions of the massless Klein-Gordon equation.
A photon can be simulated by a wave packet of electromagnetic fields with a distribution $F^{(K)}_{\Omega_{K,0}}$ of modes peaked around the frequencies $\Omega_{K,0}$ \cite{Downes,Leonhardt}.
From the perspective of an observer at a different location, the annihilation operator for photons takes the form
\begin{eqnarray}\label{operators}
\hat{a}_{\Omega_{K,0}}(\tau_K)=\int_0^{+\infty}d\Omega_K\, e^{-i\Omega_K \tau_K} F^{(K)}
_{\Omega_{K,0}}(\Omega_K)\,\hat{a}_{\Omega_K},
\end{eqnarray}
where $\Omega_{K}$ are the physical frequencies as measured in the corresponding labs, and $\tau_{K}$ is the proper times relating the Schwarzschild coordinate time $t_{K}$ with
$\tau_{K}=\sqrt{f(r_{K})}t_{K}$.
 $K=A,B$ labels either Alice or Bob,
each observer satisfies the canonical bosonic commutation relations
$[\hat{a}_{\Omega_K},\hat{a}^{\dagger}_{\Omega_K'}]=\delta(\Omega_K-\Omega'_K)$.

If Alice at location $r_A$ and time $\tau_A$ sends a wave packet $F^{(A)}_{\Omega_{A,0}}$ to the  observer Bob, the wavepacket that propagates radially will be modified when it achieves the location $r_{B}$ at the time $\tau_{B}=\Delta\tau+\frac{\sqrt{f(r_{B})}}{\sqrt{f(r_{A})}}\tau_{A}$,
 where $f(r_{A(B)})$ is the gravitational frequency shifting factor at different heights and $\Delta\tau$ represents the propagation time of the light. The modified wave packet is denoted by $F_{\Omega_{B,0}}^{(B)}$ due to curvature effects.
The time evolution of modes takes the form $i\partial_{\tau_{K}}\phi_{\Omega_{K}}^{(u)}={\Omega_{K}}\phi_{\Omega_{K}}^{(u)}$.
   This equation defines the physical frequency $\Omega_{K}$ measured by the observer at height $r_{K}$ as $\Omega_{K}=\frac{\omega}{\sqrt{f(r_{K})}}$.
   Since $\omega$ is the frequency as measured by an observer at infinity,
   if Alice sent a sharp frequency mode with $\Omega_{A}$,
   Bob will receive a mode with frequency
     $\Omega_{B}=\sqrt{\frac{f(r_{A})}{f(r_{B})}}\Omega_{A}$,
   which is the well-known gravitational redshift effect \cite{red}.
 It is feasible to utilize the relation between the annihilation operator
 to obtain the relation between
the wave packet before and after propagation \cite{earthqs1,earthqs2},
\begin{eqnarray}\label{F1}
F^{(B)}_{\Omega_{B,0}}(\Omega_B)=\sqrt[4]{\frac{f(r_B)}{f(r_A)}}
F^{(A)}_{\Omega_{A,0}}\left(\sqrt{\frac{f(r_B)}{f(r_A)}}
\Omega_B\right),\label{wave:packet:relation}
\end{eqnarray}
From the above, the observers Alice and Bob have different peak frequencies and shapes.
These changes are due to the gravitational field of the Earth and cannot be corrected simply by linear shifting frequencies.
We can decompose the mode $\bar{a}^{\prime}$ received by Bob into the mode $\hat{a}$ prepared by Alice and the orthogonal mode $\hat{a}_{\bot}$ \cite{earthqs2,earthqs3,Rohde}
\begin{eqnarray}
\bar{a}^{\prime}=\Theta_1
\hat{a}+\sqrt{1-\Theta_1^2}\hat{a}_{\bot},\label{mode:decomposition}
\end{eqnarray}
where $\Theta_1$ is the mode overlap between the wave packet
$F^{(B)}_{\Omega_{B,0}}(\Omega_B)$ and the wave packet
$F^{(A)}_{\Omega_{A,0}}(\Omega_B)$,
 \begin{eqnarray}
\Theta_1=\int_{0}^{+\infty}d\Omega_B\,F^{(B)\star}_{\Omega_{B,0}}(\Omega_B)F^{(A)}_{\Omega_{A,0}}(\Omega_B).\label{single:photon:fidelity}
\end{eqnarray}
We can use the fidelity of the quantum channel to quantify the similarity between the information transmitted and the information received.
 For a lossy channel, the fidelity of the channel is $\mathcal{F}=|\Theta|^{2}$. For a perfect channel, one finds $\mathcal{F}=1$.
 
Then we consider the downlink process,
  Bob at the satellite sends its wave packet to Alice on the Earth.
  The frequency and frequency distribution of the final returned mode are
 \begin{equation}\label{frequency1}
     \Omega_{A'}=\sqrt{\frac{f(r_{B})}{f(r_{A})}}\Omega_{B}=\Omega_{A},
   \end{equation}
   and
   \begin{eqnarray}\label{F2}
F^{(A')}_{\Omega_{A',0}}(\Omega_A)=\sqrt[4]{\frac{f(r_A)}{f(r_B)}}
F^{(B)}_{\Omega_{B,0}}\left(\sqrt{\frac{f(r_A)}{f(r_B)}}
\Omega_A\right).\label{wave:packet:relation}
\end{eqnarray}
Therefore, the wave packets overlap in the downward propagation process  is
\begin{eqnarray}
\Theta_{2}=\int_{0}^{+\infty}d\Omega_A\,F^{(A')\star}_{\Omega_{A',0}}(\Omega_A)F^{(B)}_{\Omega_{B,0}}(\Omega_A).\label{single:photon:fidelity1}
\end{eqnarray}

It is assumed that the wave packet is a normalized Gaussian wave packet
\begin{eqnarray}\label{F2}
F_{\Omega_0}(\Omega)=\frac{1}{\sqrt[4]{2\pi\sigma^2}}e^{-\frac{(\Omega-\Omega_0)^2}{4\sigma^2}},\label{packet}
\end{eqnarray}
with wave packet width $\sigma$.
The wave packet overlap $\Theta$ is obtained by using Eq. \eqref{F1} and Eq. \eqref{F2} \cite{earthqs1}
\begin{eqnarray} \label{theta}
\Theta_{1(2)}=\sqrt{\frac{2(1\pm\delta)}{1+(1\pm\delta)^2}}e^{-\frac{\delta^2\Omega_{B,0}^2}{4(1+(1\pm\delta)^2)\sigma^2}}\label{final:result},
\end{eqnarray}
where the  parameter $\delta=\sqrt[4]{\frac{f(r_A)}{f(r_B)}}-1=\sqrt{\frac{\Omega_{B}}{\Omega_{A}}}-1$ and signs $\pm$ occur for $r_B<r_A$ or $r_B>r_A$ respectively.
The explicit expression of the frequency ratio for the photon propagated between Alice and Bob has been introduced in Refs. \cite{earthqs4}.
We gain the perturbation expression of $\delta$
\begin{eqnarray}\label{bw}
\nonumber \delta&=&\delta_{Sch}+\delta_{rot}+\delta_h\\
&=&\frac{1}{8}\frac{r_S}{r_A}\big(\frac{r_A-2R}{r_A+R} \big)-\frac{(r_A\omega)^2}{4}-\frac{(r_A\omega)^2}{4}\big(\frac{3}{4}\frac{r_S}{r_A}-\frac{2r_Sa}{\omega r_A^3}\big),
\end{eqnarray}
where $\delta_{Sch}$, $\delta_{rot}$, and $\delta_h$
represent the first order Schwarzschild term, the lowest order rotation term, and
the higher order correction term, respectively.
The parameter $R$ is the height difference between Bob and Alice,
and $\omega$ denotes the Earth's equatorial angular velocity.
 If the target is located at the height $R\simeq\frac{r_{A}}{2}$, the received photon frequency at this height will not experience any frequency shift.
The wave packet overlap parameter is $\Theta_2=1-\frac{\delta^{2}\Omega_{B,0}^{2}}{8\sigma^{2}}$ in the regime $\delta\ll(\frac{\delta\Omega_{B,0}}{\sigma})^{2}\ll1$,
  which occurs for typical communication where $\Omega_{B,0}=700 \textrm{THz}$ and Gaussian bandwidth ${\sigma}=1\textrm{MHz}$  \cite{700THZ}.
Accordingly, both the final state and the wave packets overlap $\Theta_2$ are related to the range $R$ of the target.

\section{Quantum target ranging in curved spacetime}\label{section3}

\begin{figure}[ht]
\centering
\includegraphics[width=0.58\textwidth]{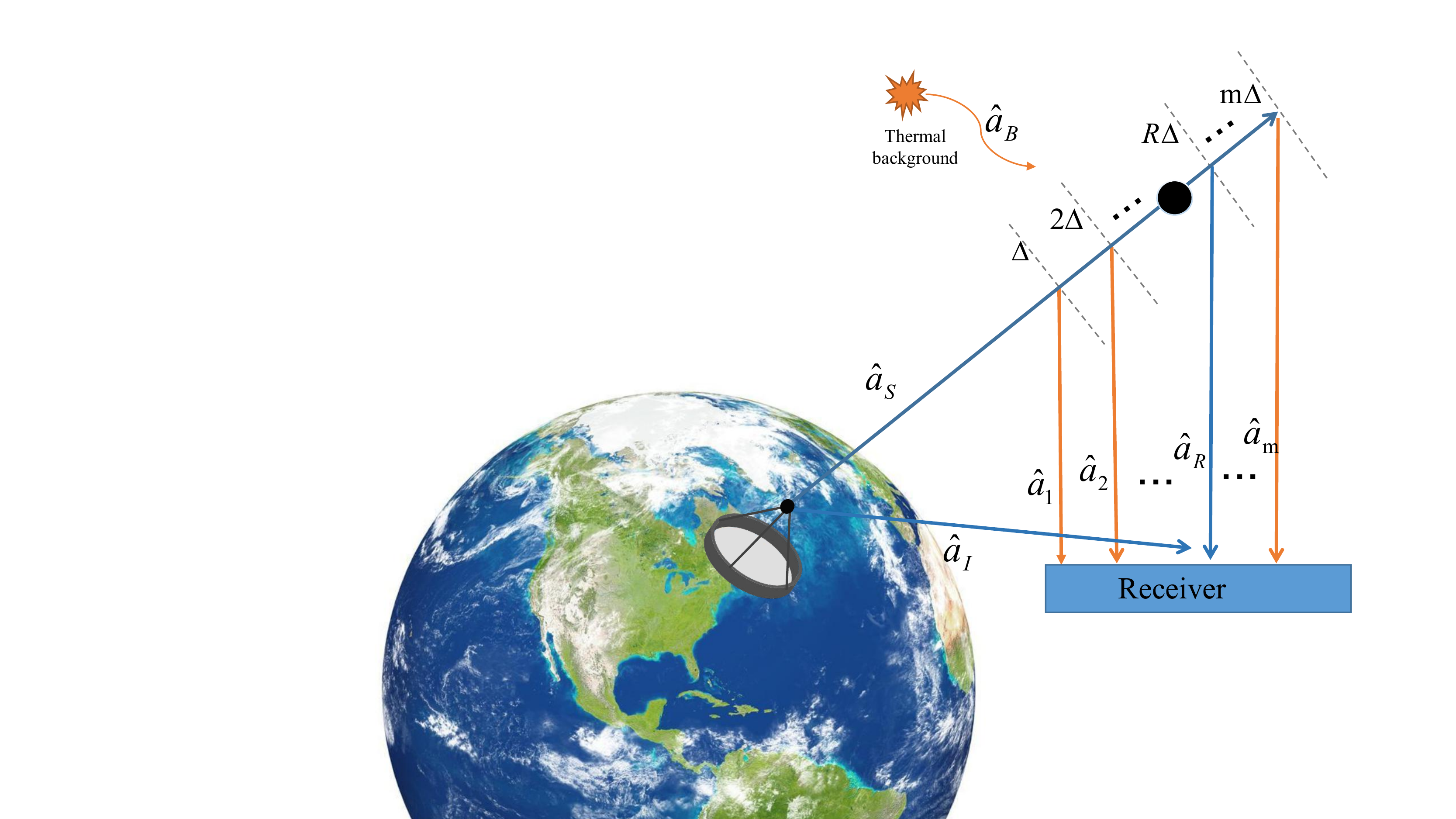}
\caption{(Color online) The entanglement-assisted ranging protocol under the
action of the Earth's gravitational field.
 The signal and idler modes are prepared from the two mode squeezed vacuum (TMSV) state.
 The signal $\hat{a}_{S}$ irradiates a spacetime target region with transmissivity $\eta$, the stored idler mode $\hat{a}_{I}$  to perform a joint quantum measurement with the reflected signals $\hat{a}_{R}$.
If the mode $\hat{a}_{R}$ is detected, the target is  $R\Delta$ away from the observer.
The target is hidden in a bright thermal noise bath, regardless of whether the target is detected or not, there will be a contribution of the thermal photons in the return signal.
}\label{fig1}
\end{figure}

The  diagram of the proposed near-Earth quantum ranging  proposal is shown Fig. (\ref{fig1}). Unlike the QI scheme, the tasks of quantum ranging can not only determine the existence of a target, but also the  distance between the observer and the target along the line of sight  \cite{ranging2,ranging1}. As shown in Fig. (\ref{fig1}),
 a low transmittance $\eta$ object is embedded in the thermal background, and
we assume that  the distance between the object and the observer can be divided  into $m\geq2$ discrete intervals.
We now have $m$ hypotheses, each corresponding to one of the $m$ range slices.
 The observer on the ground transmit a signal pulse $\hat{a}_S$ to a target region and continuously collect the returned photons $\{\hat{a}_{l}\}_{l=1}^{m}$ at the receiver side to determine the distance.
  The returned signal photon reaches the receiving side after time $\{t_{l}\}_{l=1}^{m} = 2 l \Delta/c$, where $\Delta$ is the precision of the ranging task.

In the present model,  the Earth's spacetime curvature  bring changes to the photon transmission, i.e., the frequency and shape of signal photons are affected.
In hypothesis $R$, the target is located on the distance with $R\Delta$, and the reflected mode arrives at the receiver after time $t_R=2R\Delta/c$.
Then  the problem of ranging is modeled as the determination of the reflected mode $\hat{a}_{R}$ among the continuously collect modes.
It is interesting that in the whole process (upwards plus downwards), the effects resulting from the gravitational red-shift and the gravitational blueshift on the illuminating signal $\hat{a}_{S}$ can be canceled with each other, but the gravity has the effects on the signal photons which are transmitted from the target region the receiver. The reflected mode annihilation operator at time $t_{R}$ is
\begin{eqnarray}\label{return1}
 \hat{a}_{R}=\sqrt{\eta}\hat{a}_S+\sqrt{1-\eta}({\Theta_2} \hat{a}_B+\sqrt{1-{\Theta_2}^{2}} \hat{a}_{B\bot}).
 \end{eqnarray}
where $\hat{a}_{B}$ is the thermal state with average photon number $N_{B}\gg1$.
In the flat space limit, the wave packets overlap  $\Theta_2=1$ is attained, which yeilds  $\hat{a}_{R}=\sqrt{\eta}\hat{a}_{S}+\sqrt{1-\eta}\hat{a}_{B} $ \cite{Tan,ranging1}.
The overall  transmissivity $\eta$ can provide the ratio between received power and transmitted power \cite{Zhuang4}
 \begin{equation}
\eta=\frac{P_{R}}{P_{T}}=\frac{G F^{4} A_{R} \sigma'}{(4 \pi)^{2} R^{4}}.
\end{equation}
where $\sigma'$ and $G$ are the cross-section of the target and gain of the transmit antennas, and $R$ is the transmitter-target range, respectively.
In the short range scenario, It is assumed that the form factor is unity and have an ideal pencil beam, so that its solid angle is exactly subtended through the target's cross section $\sigma'$ \cite{Zhuang4}.
This means that gain can be technically given by $G=\frac{4\pi R^{2}}{\sigma'}$.
Then the overall transmissivity is able to be
\begin{equation}
\eta=\frac{ A_{R} }{(4 \pi R)^{2}}.
\end{equation}
Through fixing the effective receive antenna collecting areas $A_{R}=0.1\textrm{m}^{2}$, we can obtain a correspondence between transmissivity $\eta$ and range $R$.
If the returned signal does not reach the receiver side at time $t_{R}$, the collected photon $\hat{a}_{l\neq R}$   is in a thermal state
\begin{equation}\label{return2}
  \hat{a}_{l\neq R}={\Theta_2} \hat{a}_B+\sqrt{1-{\Theta_2}^{2}} \hat{a}_{B\bot}.
\end{equation}


  For the entanglement-based ranging protocol, we prepare an idler-signal photon pair, where one part is emitted to the space target region as the signal photon $\hat{a}_{S}$, and the other part is retained in the local laboratory as the idler signal $\hat{a}_{I}$. The wave function of the TMSV state is
\begin{eqnarray}\label{IS-state}
|\phi^{\textrm{TMSV}}\rangle_{\mathrm{SI}}=\sum^\infty_{n=0}\sqrt{\frac{N^n_S}{(N_S+1)^{n+1}}}|n\rangle_\mathrm{S}|n\rangle_\mathrm{I},
\end{eqnarray}
where $N_S$ is the average photon number per mode.
 In the phase space representation,
$|\phi^{\textrm{TMSV}}\rangle_{\mathrm{SI}}$ is a zero mean Gaussian state whose corresponding covariance matrix $\Lambda_\mathrm{SI}=\langle[X_{1},P_{1},X_{2},P_{2}]^{T}[X_{1},P_{1},X_{2},P_{2}]\rangle$
denoted by \cite{Weedbrook}
\begin{eqnarray}\label{IS-CM}
\Lambda_\mathrm{SI}=\left(
\begin{array}{cccc}
  (2N_S+1)\textbf{I}_{2} & 2S_{p}\textbf{Z}_{2}  \\
  2S_{p}\textbf{Z}_{2} & (2N_S+1)\textbf{I}_{2}
\end{array}
\right),
\end{eqnarray}
where $X_{1}=\frac{1}{\sqrt{2}}(\hat{a}_{S}+\hat{a}_{S}^{\dag})$, $P_{1}=\frac{1}{i\sqrt{2}}(\hat{a}_{S}-\hat{a}_{S}^{\dag})$, $X_{2}=\frac{1}{\sqrt{2}}(\hat{a}_{I}+\hat{a}_{I}^{\dag})$, and $P_{2}=\frac{1}{i\sqrt{2}}(\hat{a}_{I}-\hat{a}_{I}^{\dag})$.
With $S_{p}=\sqrt{N_S(N_S+1)}$, $\mathbf{I}_{2}$ and $\mathbf{Z}_{2}$ are the identity matrix and Pauli matrix.

 In the entangled ranging protocol, we assume that the target position is on the center of the $R\Delta$ slice.
 Then the overall return-idle state by the receiver side is \cite{ranging1,ranging2}
 \begin{equation}\label{E-input}
\hat{\rho}_{R}^{E}=\left(\otimes_{l \neq R} \hat{\sigma}_{\hat{a}_{l}}^{(B)}\right) \otimes \hat{\Xi}_{\hat{a}_{R} \hat{a}_{I}}^{(T)},
\end{equation}
where $\hat{\sigma}_{\hat{a}_{l}}^{(B)}$ is a set of $M$ mode thermal photon signals, and the average number of photons of each thermal photon signal is $N_{B}$.
Moreover, $\hat{\Xi}_{\hat{a}_{R} \hat{a}_{I}}^{(T)}$
is the joint measurement state of $M$ mode signal-idler photon pairs returned at $R \Delta$, where the returned mode is Eq. (\ref{return1}).
Each  pair in the state is described by  the covariance matrix
 \begin{eqnarray}\label{CM3}
\Lambda'_\mathrm{SI}=\left(
\begin{array}{cccc}
  (1+2\eta N_S+2\Theta_2^{2}N_B)\textbf{I}_{2} & 2\sqrt{\eta}S_{p}\textbf{Z}_{2}  \\
  2\sqrt{\eta}S_{p}\textbf{Z}_{2} & (2N_S+1)\textbf{I}_{2}
\end{array}
\right).
\end{eqnarray}

   In the classical strategy, the position of the target is on the center of the $R\Delta$ slice, and the state of the output signal at the receiving side can be written as
 \begin{equation}
\hat{\rho}_{R}^{C}=\left(\otimes_{l \neq R} \hat{\sigma}_{\hat{a}_{l}}^{(B)}\right) \otimes \hat{\sigma}_{\hat{a}_{R}}^{(T)}.
\end{equation}
  The target state $\hat{\sigma}_{\hat{a}_{R}}^{(T)}$ is the $M$ mode reflect signal embedded in a thermal noise background, which is generated through the thermal loss channel Eq. (\ref{return1}). The covariance matrix of the target state is $[1+2\Theta_2^{2}N_{B}]\mathbf{I}_{2}$.

The performance of quantum ranging strategy is measured by the bound of error probability of the hypotheses.
 In both of these detection schemes, the input state $\hat{a}_{S}$ is assumed to have a positive P function.
 For mixed states, it is difficult to obtain a general bound for the Helstrom limit $P_{H}\left(\left\{\rho_{n}, p_{n}\right\}\right)$.
 An upper bound can be obtained from the pretty good measurement (PGM) \cite{ranging2,PGM1,PGM2,QCD4} described by  the positive operator-valued measure.
 The error probability is given by
\begin{equation}
P_{E}^{\mathrm{PGM}}=1-\sum_{n=0}^{m-1} p_{n} \operatorname{tr}\left(\Pi_{n}^{\mathrm{PGM}} \rho_{n}\right) \geq P_{H}\left(\left\{\rho_{n}, p_{n}\right\}\right).
\end{equation}
Fortunately, the form of error probability of ranging protocol based on the fidelity is easier to calculate \cite{ranging2,QCD4,F}
\begin{equation}
P_{H} \leq P_{H, U B}:=2 \sum_{n^{\prime}>n} \sqrt{p_{n^{\prime}} p_{n}} F\left(\rho_{n^{\prime}}, \rho_{n}\right)
\end{equation}
and
\begin{equation}
P_{H} \geq P_{H, L B}:=\sum_{n^{\prime}>n} p_{n^{\prime}} p_{n} F^{2}\left(\rho_{n^{\prime}}, \rho_{n}\right),
\end{equation}
where $\emph{F}$ is the Bures' fidelity
\begin{equation}
F(\rho, \sigma):=\|\sqrt{\rho} \sqrt{\sigma}\|_{1}=\operatorname{tr} \sqrt{\sqrt{\rho} \sigma \sqrt{\rho}}.
\end{equation}
It is assumed that all slice may occur with is equal probability, so that $p_{n}=\frac{1}{m}$ for any $n$. Then we get the simplified boundary
\begin{equation}
P_{H, U B}:=(m-1) F^{M}\left(\rho_{n^{\prime}}, \rho_{n}\right),
\end{equation}
\begin{equation}
P_{H, L B}:=\frac{m-1}{2 m} F^{2M}\left(\rho_{n^{\prime}}, \rho_{n}\right).
\end{equation}

Combining the symmetry of the whole slice range and the fidelity of the classical strategy, the lower bound of the optimal performance of the classical detection strategy can be calculated by the positive P function input state \cite{ranging1}.
The influence of the gravitational field of the Earth on the lower bound of classical ranging strategy is 
\begin{eqnarray}\label{bound1}
\nonumber P'_{C, \mathrm{LB}}&=&\frac{m-1}{2 m}F^{2M}(\sigma^{(T)},\sigma^{(B)})\\
&\simeq&\frac{m-1}{2 m} \exp \left[-\frac{2 M \eta N_{S}}{1+2\Theta_2^{2} N_{B}}\right],
\end{eqnarray}
where $\Theta_2$ is the mode overlap of the  loss channel induced by the Earth's spacetime effect given in Eq. (\ref{mode:decomposition}).

In the quantum ranging protocol, conditioned on the target range being $R\Delta$, the overall state at the receiver is Eq. (\ref{E-input}).
Due to the symmetry
and the structure of the overall return-idler states,
the error probability of the multiple hypothesis testing problem is equal to the error exponent of discriminating two three-mode zero-mean Gaussian states $(\hat{\sigma}_{\hat{a}_{1}}^{(B)} \otimes \hat{\Xi}_{\hat{a}_{2} \hat{a}_{I}}^{(T)},  \hat{\Xi}_{\hat{a}_{1} \hat{a}_{I}}^{(T)}\otimes \hat{\sigma}_{\hat{a}_{2}}^{(B)})$.
The returned idler modes under two different quantum ranging hypotheses are distinguished
 \begin{equation}
\Lambda^{(1)'}_{12I}=\left(\begin{array}{ccc}
(1+2\eta N_S+2\Theta_2^{2}N_B)\textbf{I}_{2} & \mathbf{0} & 2 \sqrt{\eta} S_{p} \mathbf{Z}_{2} \\
\mathbf{0} & \left(2 N_{B}+1\right) \mathbf{I}_{2} & \mathbf{0} \\
2 \sqrt{\eta} S_{p} \mathbf{Z}_{2} & \mathbf{0} & \left(2 N_{S}+1\right) \mathbf{I}_{2}
\end{array}\right),
\end{equation}
\begin{equation}
\Lambda^{(2)'}_{12I}=\left(\begin{array}{ccc}
\left(2 N_{B}+1\right) \mathbf{I}_{2} & \mathbf{0} & \mathbf{0} \\
\mathbf{0} & (1+2\eta N_S+2\Theta_2^{2}N_B)\textbf{I}_{2} & 2 \sqrt{\eta} S_{p} \mathbf{Z}_{2} \\
\mathbf{0} & 2 \sqrt{\eta} S_{p} \mathbf{Z}_{2} & \left(2 N_{S}+1\right) \mathbf{I}_{2}
\end{array}\right).
\end{equation}
We can also attain the lower bound \cite{fidelity} for the error probability of the entanglement enhanced ranging protocol in the curved spacetime,
\begin{eqnarray}\label{bound3}
\nonumber P'_{E, \mathrm{LB}}&=&\frac{m-1}{2 m}F^{2M}(\Lambda^{(1)'}_{12I},\Lambda^{(2)'}_{12I})\\
&\simeq&\frac{m-1}{2 m} \exp \left[-\frac{2 M \eta N_{S}}{1+\Theta_2^{2} N_{B}}\right].
\end{eqnarray}
The corresponding bounds $P_{C, \mathrm{LB}}$ and $P_{E, \mathrm{LB}}$ in the flat space can be obtained through setting $\Theta_2=1$ in the Eqs. (\ref{bound1}-\ref{bound3}). In the curved spacetime, the wave packet overlap parameter   $\Theta_2=1-\frac{\delta^{2}\Omega_{B,0}^{2}}{8\sigma^{2}}$ is less than 1.
 Therefore, we can conclude that the spacetime effects of the Earth can reduce the error probability of ranging tasks.
 This result is similar to the performance of QI  in the gravitational field of the Earth \cite{QIqianqian}.

Then we give a physical interpretation for this phenomenon.
When the quantum ranging task operates in a curved spacetime background,
 both the signal  $\hat{a}_{S}$  and
  the thermal photon  $\hat{a}_{B}$ are deformed due to the gravitational effects. 
In the present scheme, the gravitational redshift effect on the initial signal emitted by Alice is designed to be eliminated by an opposite gravitational blueshift factor $\sqrt{\frac{f(r_{B})}{f(r_{A})}}$, since the signal will be sent downward from the satellite to Earth.
 The  gravitational effects are found to influence the fidelity of the quantum channel between Alice and the final state.
The mode overlap $\Theta$ between the final state
$F^{(A')}_{\Omega_{A',0}}$ and the wave packet
$F^{(A)}_{\Omega_{A,0}}$ is
 \begin{eqnarray}
\Theta=\int_{0}^{+\infty}d\Omega_A\,F^{(A')\star}_{\Omega_{A',0}}(\Omega_A)F^{(A)}_{\Omega_{A,0}}(\Omega_A).\label{single:photon:fidelity}
\end{eqnarray}
which describes the fidelity of the channel between Alice and the final state. One finds $F^{(A')}_{\Omega_{A',0}}(\Omega_A)=F^{(A)}
_{\Omega_{A,0}}(\Omega_A)$ and $\Theta=\int_{0}^{+\infty}d\Omega_A\,| F^{(A)}_{\Omega_{A,0}}(\Omega_A)|^{2}=1$.
 However,
the thermal photons $\hat{a}_{B}$ returned from the target region  is only affected by  the gravitational  blue-shift because they only experience a one-way  transmission process. When the returned signal does not reach the receiver side at time $t_{R}$, the collected photon $\hat{a}_{l\neq R}$ is changed from $\hat{a}_B$ to ${\Theta_2}\hat{a}_B+\sqrt{1-\Theta_2^{2}}\hat{a}_{B\perp}$,
which demonstrate the spacetime effects decrease the
contribution of the thermal background to the returned signal.

Here the quantum protocol outperforms the optimal classical protocol for error probability by using the bounds derived from the above calculation.
We are interested in how the spacetime effect of the Earth affect the performance of the quantum ranging tasks.
We figure out the maximum potential  advantage of the quantum enhanced ranging protocol in the flat space and the Earth spacetime, respectively.
 The maximum potential quantum advantage can be defined by the difference between the classical and quantum lower bounds \cite{thermal}
 \begin{eqnarray}\label{max}
 f\textrm{-}\Delta P_{\textmd{max}}&=&P_{C, \mathrm{LB}}-P_{E, \mathrm{LB}},\\
  c\textrm{-}\Delta P_{\textmd{max}}&=&P'_{C, \mathrm{LB}}-P'_{E, \mathrm{LB}},
 \end{eqnarray}
  where $f$-$\Delta P_{\textmd{max}}$ to represent the maximum potential advantage of quantum ranging protocol in the flat spacetime.  Similarly, $c$-$\Delta P_{\textmd{max}}$ denotes the maximum potential advantage of quantum ranging protocol in the near-Earth curved spacetime.
The difference $\textrm{D}(\Delta P_{\textmd{max}})$ between the maximum potential quantum advantage in the curved spacetime and its flat counterpart can be denoted by
\begin{equation}\label{Dmax}
  \textrm{D}(\Delta P_{\textmd{max}})=(c\textrm{-}\Delta P_{\textmd{max}})-(f\textrm{-}\Delta P_{\textmd{max}}).
 \end{equation}

\begin{figure}[ht]
\centering
\includegraphics[width=0.495\textwidth]{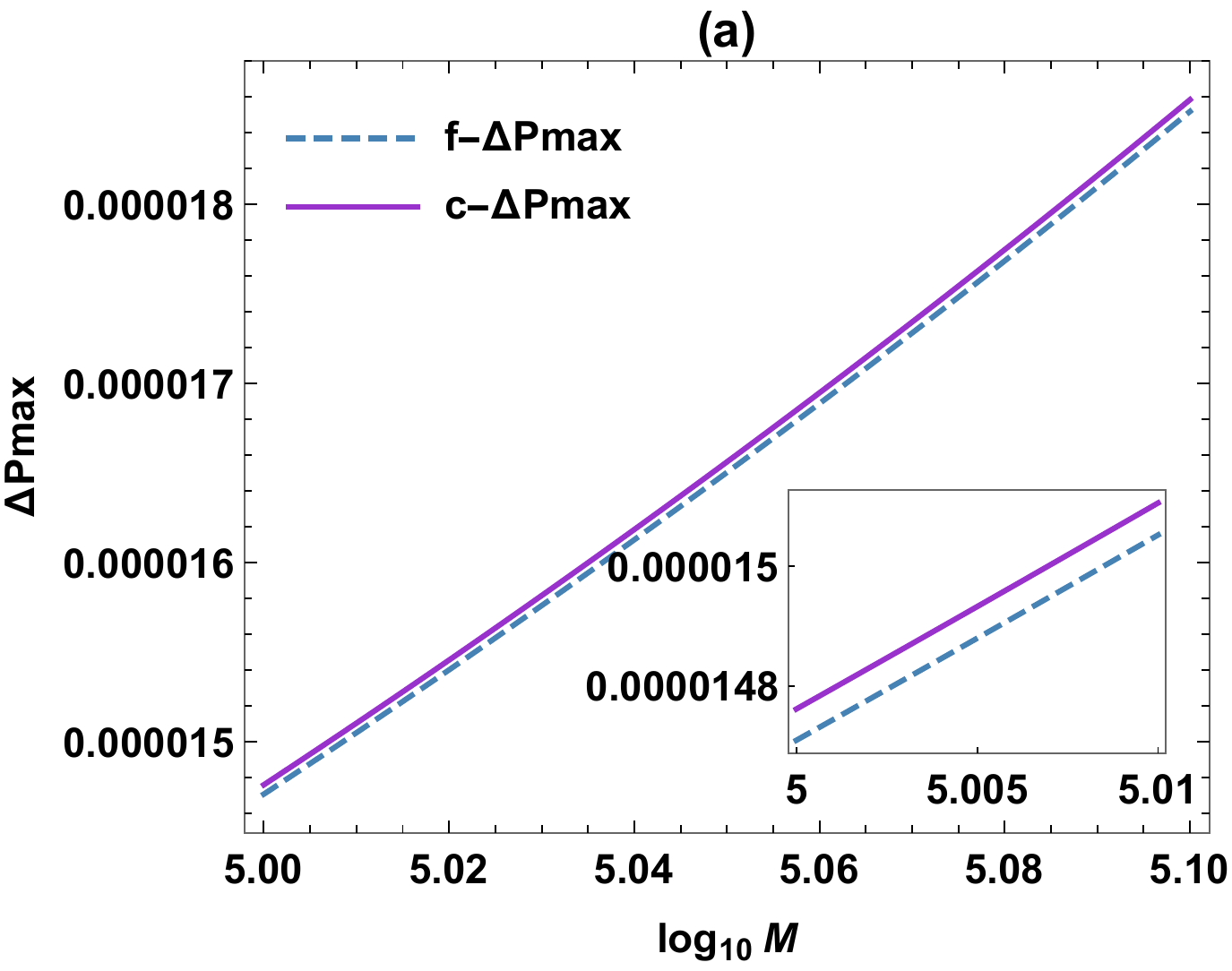}
\includegraphics[width=0.495\textwidth]{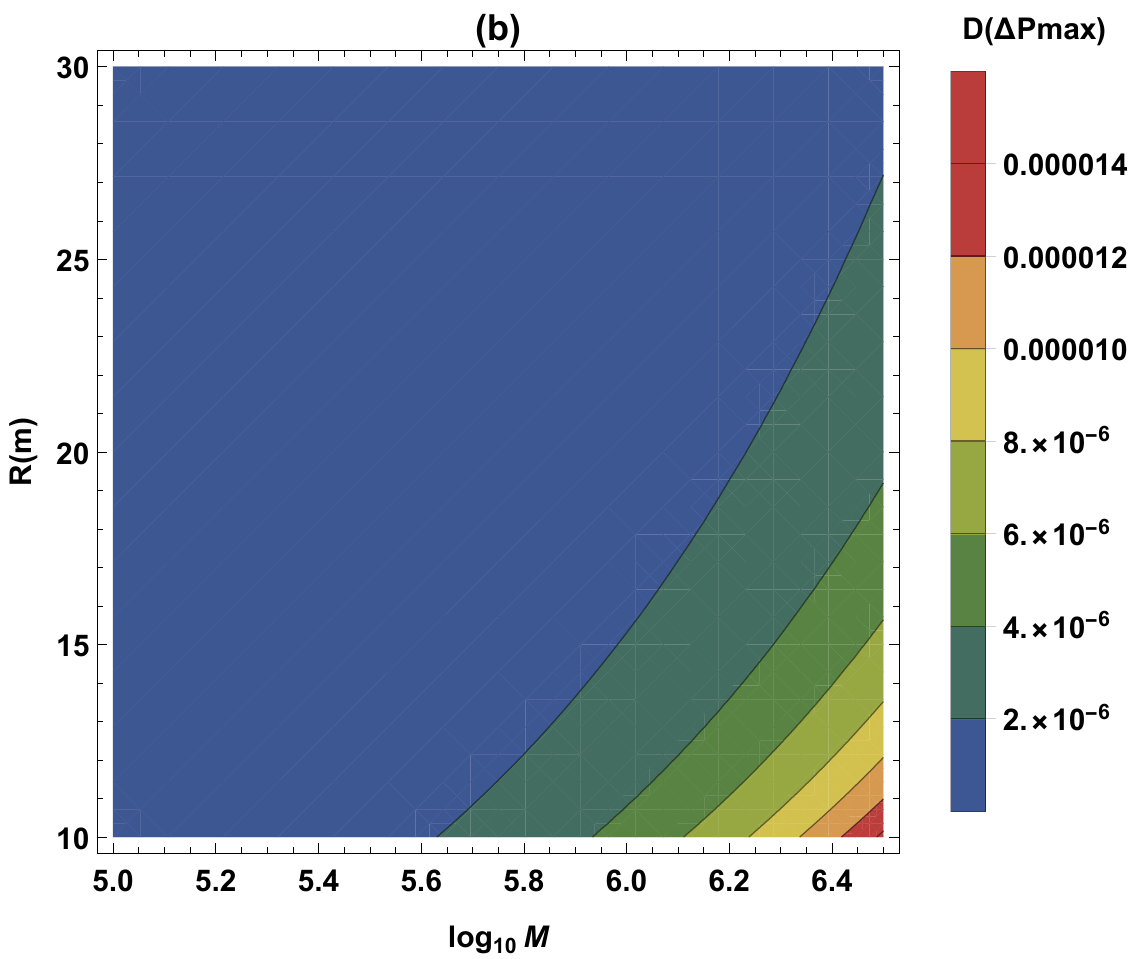}
\caption{(Color online) (a) The maximum potential advantage of quantum ranging strategy in the near-Earth spacetime versus the flat spacetime counterpart. The transmitter-target range is fixed as  $R=30$m.
(b) The difference of the maximum quantum potential advantage  in curved spacetime and the flat spacetime case. Here the number of range slices $m=10$, the signal brightness $N_S=0.01$ and the environmental noise $N_B=20$.}\label{figM}
\end{figure}

Fig. \ref{figM}(a) shows the maximum potential quantum advantage with respect to the number of copies $M$ in different spacetime backgrounds.
It is shown that the maximum potential advantage of quantum ranging protocol increases with the growth of the number of copies $M$, either in the curved spacetime or in the flat spacetime.
As mentioned above, the maximum potential quantum advantage in the curved spacetime is higher than the flat spacetime.
In other words, the existence of gravity promotes  the maximum possible advantage for the quantum ranging strategy.
In Fig. \ref{figM}(b), we plot the difference D($\Delta P_{\textmd{max}}$) between the maximum potential quantum advantage in the curved spacetime and its flat counterpart. The results show that the
D($\Delta P_{\textmd{max}}$) is always positive, which indicates that the maximum potential quantum advantage in the curved spacetime is higher than that in flat spacetime. Furthermore, the difference increase as the copies of the transmitted modes $M$.
It is concluded that increasing the  copies of transmitted mode can promote the maximum potential advantage of the quantum ranging tasks.

\begin{figure}[ht]
\centering
\includegraphics[width=0.52\textwidth]{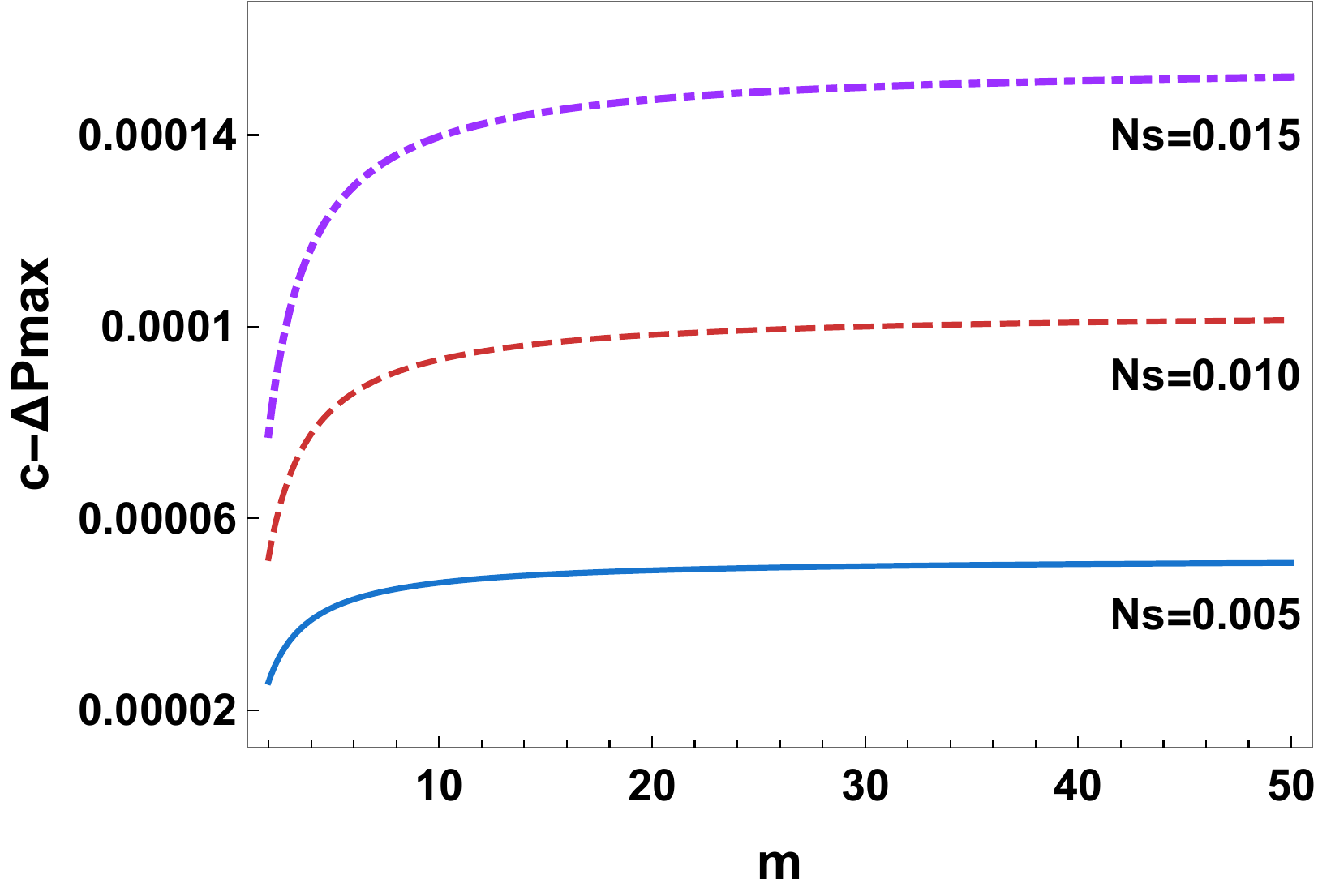}
\caption{(Color online) The maximum potential advantage of quantum ranging strategy in near-Earth spacetime versus the number of range slices $m$. The environmental thermal background noise $N_{B} = 20$, the transmitter-target range $R = 30$m and the number of modes $M = 10 ^ {5.8}$. }\label{figm}
\end{figure}

 The quantum ranging scheme divides the range between the ground observer and the target into $m\geq2$ length slices, where each hypothesis corresponds to the target being in one of the slices.
In Fig. \ref{figM}(b), the maximum potential quantum advantage in curved spacetime surpass the flat spacetime.
To better understand the gravitational field of Earth on the performance of quantum ranging. Fig. (\ref{figm}) shows the maximum potential quantum advantage in curved spacetime as a function of the number of distance slices $m$ for different number of emitted signal photons $N_{S}$. It can be seen that the maximum potential advantage of quantum strategy increases quickly for some small  number of range slices in curved spacetime. Nonetheless, $c-\Delta p_{\textmd{err}}^{\textmd{max}}$ can not  been raised sharply through dividing the range to more slices.
In addition,  by increasing the emitted signal photons for $N_{S}\ll1$ can enhance the maximum potential advantage of quantum strategy in the curved spacetime.  We can choose the appropriate signal energies and number of range slices to obtain the maximum potential advantage for the quantum ranging protocol in the near-Earth curved spacetime.

\section{Conclusions} \label{section4}

 In this paper, we suggest a near-Earth quantum ranging protocol to detect the distance between the observer and the target, in which the binary quantum hypothesis in QI is replaced by the $m$ hypothesis. In the present proposal, the performance of quantum ranging is affected by the Earth's spacetime curvature since the wave packet overlap and  the overall transmissivity of the photons are deformed by the spacetime effects.
  It is shown that  the maximum potential advantage of a quantum strategy in curved spacetime has distinct advantages over its counterpart in flat spacetime.  This indicates the spacetime curvature  can reduce the error probability of quantum ranging tasks.  This is because the effect of the gravitational red-shift and blue-shift on the  entangled signal beam can  cancel each other, while the gravity always affects the spatial returned thermal mode.  Moreover, It is found that the  copies of transmitted modes can promote the maximum potential advantage of the quantum ranging tasks.
  The maximum potential advantage of quantum strategy increases quickly for some small number of range slices.
  Therefore, one can choose the appropriate signal energies and the number of range slices to obtain a better maximum potential advantage for the quantum ranging task in the near-Earth curved spacetime.

\begin{acknowledgments}
This work is supported by the National Natural Science Foundation of China under Grant No. 12122504 and  No.11875025.
\end{acknowledgments}



\end{document}